On the Hall Effect in the pseudogap phase of cuprates

Lev P. Gor'kov
*NHMFL, Florida State University, 1800 East Paul Dirac Drive, Tallahassee Florida 32310, USA
and L.D. Landau Institute for Theoretical Physics of the RAS, Chernogolovka 142432, Russia*

Gregory B. Teitel'baum
*E.K. Zavoiskii Institute for Technical Physics of the RAS, Kazan 420029, Russia*

(Dated     )

Angle resolved photoemission (ARPES) experiments reveal the coherent electronic excitation in the energy spectrum of under- and optimally doped cuprates at temperatures above their temperature of transition into superconducting state only near nodal points on the parts of the "bare" Fermi surface known as the Fermi arcs. That raises questions about the meaning and possible interpretations of the experimental data. Below we derive expressions for the Hall coefficient for the Fermi arc model both in weak and strong magnetic fields.

Understanding high temperature superconductivity (HTSC) in cuprates is the pressing issue both on the theory side of the problem and for practical implications. While in superconductors of "old" generation it was at least known that superconductivity emerges from the normal Landau Fermi liquid phase, HTSC is preceded by the so-called pseudogap phase with many abnormal properties. Indeed, the most unexpected feature revealed in the angle resolved photoemission (ARPES) experiments is that the coherent electronic excitations exist only on the Fermi arcs along the "bare" Fermi surface and separated from each other by large energy gaps [1].

It then is absolutely unclear how to interpret even such basic experimental data as for resistivity or the Hall coefficient. For the first time the importance of the question was probably realized when it turned out that the Hall coefficient [2] in $La_{1-x}Sr_xCuO_4$ (LSCO) manifests the activation temperature dependence [3].

So far, ARPES experiments were performed on the two-layer $Bi_2Sr_2CaCu_2O_8$ (Bi2212), single-layer Bi2201 and LSCO (see recent review [4]). As yet, ARPES is not available for Hg1201 and YBCO because of problems with the surface [5]. The consensus is that, except for minor details, ARPES findings bear the general character and reflect the basic physics. It is all the more remarkable that the recent experiments [6] added new and convincing evidence in support of the Fermi arcs concept.

In brief, it was found in [6] that characteristic for temperature dependence of resistivity of the clean Hg1201 (and few others cuprate compounds) is the contribution proportional to square of the temperature, as if in a Fermi liquid. In [6] the latter was related to free carriers on the Fermi arcs. A rigorous model of carriers on the Fermi arcs interacting via short-range interactions has been carefully examined in [7]. (For convenience, some results [7] will be written out below).

In its most general form the kinetic equation is:

$$\frac{dn}{dt} = I_{col}. \qquad (1)$$

On the left, the total derivative $(dn/dt)$ accounts for variations of the Fermi distribution of the quasiparticles at their motion in the real and in the momentum space in the presence of external fields. The right hand side term, the collision integral $I_{col}$ is responsible for the relaxation processes.

In weak constant electric and magnetic fields the kinetic equation (1) is presented in the commonly-accepted form:

$$e(\vec{E} + \frac{1}{c}[\vec{v} \times \vec{H}])\vec{\nabla}_{\vec{p}} n(\vec{p}) = I_{coll}. \qquad (2)$$

Changes $n_1$ of the Fermi distribution function $n(\vec{p})$ are also small. At a non-zero current the system of charged carriers moves as a whole with a drift velocity $\vec{u}$. Correspondingly, the left hand side in (2) reduces to:

$$e\{(\vec{E} \cdot (\vec{v} - \vec{u})) - \frac{1}{c}([\vec{v} \times \vec{H}] \cdot \vec{u})\} \frac{\partial n_o}{\partial \varepsilon}, \qquad (3)$$

(Here $(\partial n_o / \partial \varepsilon)$ is the derivative of the equilibrium Fermi function; $\vec{v} = d\varepsilon(\vec{p})/d\vec{p}$). In the above approximation the expressions for the quadratic resistivity term and for the Hall coefficient, correspondingly, are [7]:

$$\rho_{2D}(T) = \left(\frac{\pi\hbar}{e^2}\right) \frac{4\pi^2 |Z\tilde{V}(1;2)|^2}{3(\Delta\varphi)^2} \sqrt{\frac{\Delta}{\varepsilon_F^*}} \left(\frac{T}{\varepsilon_F^*}\right)^2 \qquad (4)$$

and

$$R_H = \frac{1}{ecn_{eff}} > 0. \qquad (5)$$

In Eq.(4) $Z\tilde{V}(1;2)$ is the renormalized matrix element of the electron-electron interaction, $m^* = Z^{-1}m$ is the renormalized mass,. The combination $\varepsilon_F^* = p_F^2 / 2m^*$ is defined as the renormalized Fermi energy. In Eq. (16) $\Delta = v_F p_F[(K/4p_F) - 1]$ $K^{\|}$ is projection of the Umklapp vector $\vec{K} = (2\pi/a, 2\pi/a)$ on the diagonal in the Brillouin zone (BZ); $\Delta\varphi$ is the arc width and $n_{eff} = (\Delta\varphi) p_F^2 / \pi^2$ gives the effective number of carriers. (Each of these parameters can be found experimentally).

Rigorously speaking, expression (5) for the Hall coefficient applies only in the weak field regime $\omega_c \tau_{col} \ll 1$ ($\omega_c = (eH/m^*c)$ [7]. (Notation $\tau_{col}$ stands for the mean scattering time). Meanwhile, by the order of magnitude $\tau_{col}$ typically is: $\tau_{col} \propto \varepsilon_F / T^2$, i.e., $\tau_{col}$ is large in clean samples. For instance, at $T = 100\ K$ the magnetic field should be much smaller than $3-5\ Tesla$ for (5) to apply. Data for $R_H$ in

Hg1201, YBCO and LSCO in the whole temperature range were obtained only in higher fields (see [2] and [8] and references in the latter).

Below we study the Hall Effect in frameworks of the Fermi arcs model in the opposite limit $\omega_c \tau_{col} \gg 1$. Unlike the situation common in most semiconductors, it turned out that the expression (5) of $R_H(T)$ remains correct in both limits weak and extremely strong magnetic fields.

Assuming the tetragonal symmetry for the $CuO_2$-plane in cuprates, the Hall coefficient can be defined as $R_H = \sigma_{xy}(\sigma_{xx}/\sigma_{yy}) \equiv \sigma_{xy}$ (below we use the system of coordinates with the $x, y$-axes along the two diagonals of the BZ). For normal metals in the limit of strong field the value of the non-diagonal conductivity component $\sigma_{xy}$ depends critically on whether electrons at the Fermi surfaces in the magnetic field move along closed or open trajectories (see e. g. [9]). In particular, in the former case:

$$\sigma_{xy} = -\frac{|e|c}{H} n . \quad (6)$$

(Here $n$ is the number of carriers for a closed Fermi surface).

Returning to Eq. (1), in strong magnetic fields $\omega_c \tau_{col} \gg 1$ the collision integrals obviously should be omitted. Finding the distribution functions for carriers placed in the strong magnetic field and a weak electric field reduces to solving the equations of motion, as given by the Liouville's Theorem:

$$\frac{dn}{dt} = 0 . \quad (7)$$

The approach to solving (7) was elaborated in [10]. (See summary of the method in [9]).

Unlike weak fields, as in Eq.(3), the charges now move along trajectories bent by the strong magnetic field. Because of that presenting the distribution functions in terms of the Cartesian momentum components, $p_x, p_y$ is not convenient. Instead, in [10] it was suggested to go over to the new variables: energy $\varepsilon$ and the time variable $\tau$ defined via the relation:

$$d\tau = \frac{ds}{v_\perp}(c/eH) \quad (8)$$

(Here $ds$ is the element of "length" along the Fermi surface: $ds^2 \equiv dp_x^2 + dp_y^2$ ; $v_\perp^2 = v_x^2 + v_y^2$). Eq. (8) expresses that in a strong field, $H$ carriers move along the Fermi surface with high speed.

For a weak *electrical* field Eq. (7) takes the form [9, 10]:

$$\frac{dn}{dt} \approx -\frac{dn_0}{d\varepsilon} e(\vec{v} \cdot \vec{E}) + \frac{\partial n_1}{\partial \tau} \approx 0 . \quad (9)$$

(The collision integrals were discarded). As to $n_1$, the latter is presented [9, 10] in the form:

$$n_1 = (\partial n_0 / \partial \varepsilon) e (\vec{E} \cdot \vec{g}). \qquad (10)$$

The equation for $\vec{g}$ is:

$$\partial \vec{g} / \partial \tau = \vec{v}. \qquad (11)$$

It is important that the solution of (11) cannot contain a constant vector $\vec{g}_0$. (Such term would result in a meaningless constant shift of the distribution function (10)). As to the further calculations, the latter for the Fermi arcs model turn out to be, in some sense, even simpler than in [10].

In fact, there is no dilemma concerning the motion along either open or closed trajectories in the momentum space: the carriers cannot go away from arcs because on both sides of the arc the distribution function is restricted by large energy gaps. In [10] the "time" $\tau$ for a carrier was defined formally, by its initial position on the Fermi surface. The physical meaning of such procedure [10] is that it indirectly characterizes carriers by their velocities $ds / d\tau$ along the Fermi surface; as follows from Eq. (8). The Fermi arcs being separated from each other by large energy gaps where the distribution function is zero, the time variables $\tau$ of Eq. (8) can be introduced for each Fermi arc.

One can now but refers to Eq. (84, 11) in [9]. The latter reads now as the sum over the four arcs:

$$\sigma_{\alpha\beta} = \frac{2e^3 H}{(2\pi)^2 c} \sum \int_{\Delta\varphi_i} v_\alpha g_\beta d\tau \equiv \frac{2e^3 H}{\pi^2 c} \int_{\Delta\varphi} v_\alpha g_\beta d\tau, \qquad (12)$$

(Sign $\Delta\varphi$ at the integral reminds that integration in (12) is limited by one Fermi arc). For definiteness, chose $\sigma_{xy}$:

$$\sigma_{xy} = \frac{2e^3 H}{\pi^2 c} \int_{\Delta\varphi} v_x g_y d\tau. \qquad (13)$$

After writing out the equation

$$\frac{d\vec{p}}{d\tau} = \frac{e}{c} [\vec{H} \times \vec{v}] \qquad (14)$$

in the $x, y$-components, solving Eq.(11) gives $g_y = -(c / eH) p_x$ while $v_x = (c / eH)(dp_y / d\tau)$; one finds that integration over $\tau$-variable in (13) is actually the integration over the momentum component $p_y$:

$$\sigma_{xy} = \frac{2e^3 H}{\pi^2 c} \int_{\Delta\varphi} v_x g_y d\tau = -\frac{2ec}{\pi^2 H} \int_{\Delta\varphi} p_x dp_y, \qquad (15)$$

where the last integral is over the area $\delta S$ limited by the single arc:

$$\sigma_{xy} = -\frac{ec}{H} n_{eff}. \qquad (16)$$

In the isotropic case $\delta S = (p_F^2/2)\Delta\varphi$ and $n_{eff} = (\Delta\varphi)p_F^2/\pi^2$ is defined as in Eq. (5). At $\Delta\varphi \Rightarrow \pi/2$ one gradually returns to the expression of the Hall constant for the closed Fermi surface [9]. In the anisotropic case the number of carriers is expressed in terms of the area $\delta S$ limited by the single arc. (For holes: $(-e) \Rightarrow |e|$. One sees from Eq.(15) that introduction of the time variable $\tau$ for the finite Fermi arc was nothing more than a convenient mathematical device).

Eqs. (15, 16) conclude the proof.

In summary, in frameworks of the Fermi arc model the Hall coefficient $R_H(T)$ determines the actual number of carriers even if the arc width is not small. In underdoped cuprates one obtains from $R_H(T)$ the value and the temperature dependence of the Fermi arc width $\Delta\varphi(T)$ provided that the positions $p_F$ of the Fermi surface at the nodal points is known. Experimentally, the latter are rather close to the points $(\pm\pi/2, \pm\pi/2)$ in the BZ [4].

Acknowledgements

The work of L.P.G. was supported by the NHMFL through NSF Grant No. DMR-1157490, the State of Florida and the U.S. Department of Energy; that of G.B.T. through the RFBR under Grant No. 10-02-01056.